\documentclass[12pt,letterpaper,dvips]{article}
\usepackage{amstex}
\usepackage{latexsym}
\usepackage{graphicx}
\usepackage{amssymb}
\usepackage{wrapfig}

\def\etal{et al.}

\def\lya{Ly$\alpha$ }

\def\kms{km~s$^{-1}$ }
\def\cm2{\, \rm cm^{-2}}

\def\smm{\sum\limits}

\def\cmma{\;\;\; ,}

\def\intl{\int\limits}

\def\cosmZ{$<{\cal Z}(Z)>$}

\def\colH{$N$(H I)}
\def\Dv{$\Delta v$}
\def\intl{\int\limits}
\def\smm{\sum\limits}
\def\apj{ApJ}
\def\apjsupp{ApJS}
\def\mnras{MNRAS}

\def\araa{ARAA}
\def\aj{AJ}

\title{ON THE EVOLUTION OF DAMPED {\lya} SYSTEMS TO GALACTIC DISKS} 

\author{ ARTHUR M. WOLFE$^1$
\& JASON X. PROCHASKA$^1$ \\
Department of Physics, and Center for Astrophysics and Space Sciences; \\
University of California, San
Diego; \\
C--0424; La Jolla; CA 92093\\}

\footnotetext[1]{Visiting Astronomer, W.M. Keck Telescope.
The Keck Observatory is a joint facility of the University
of California and the California Institute of Technology.}

\begin{document}
\maketitle

\begin{abstract} 
The mean metallicity of the thick disk of the Galaxy is 0.5 dex higher
than that of the damped {\lya} systems. This has been interpreted to argue
that stars in the former do not arise out of gas in the latter. Using
new metallicity and H I column-density data we show the metal-rich damped
systems do contain sufficient baryons at the thick-disk metallicity to account for
the stellar masses of thick disks. Comparing our kinematic data with the
metallicities we show that damped {\lya} systems exhibiting the largest profile
velocity widths, {\Dv}, span a narrow range of high metallicities, while systems with
small {\Dv} span a wider range of metallicities. This is naturally
explained by passage of the damped {\lya} sightlines through rapidly 
rotating disks with negative
radial gradients in metallicity. The systematically lower {\colH} of systems with
high {\Dv} indicates (a) the gaseous disks have centrally located holes, and (b)
an apparent inconsistency with the protogalactic clump model for damped
{\lya} systems. The higher metallicity of systems with low {\colH} further 
implies that stars rather than gas dominate the baryonic content of the most
metal-rich damped systems. \end{abstract}

Keywords: cosmology---galaxies: evolution---galaxies: 
quasars---absorption lines 

\clearpage

\section{INTRODUCTION}

The  collapse of a spheroidal protogalaxy to the centrifugally supported disk
of the Galaxy
was inferred from correlations 
between the metallicities and kinematics of old stars in the
solar neighborhood (Eggen, Lynden-Bell, and Sandage 1962).
But subsequent studies have not sorted out
the sequence of events leading to the formation of stellar populations comprising
the halo, the thick disk, and the thin disk (e.g., Majewski 1993).
The damped
{\lya} absorption systems, a population of
H I layers widely believed to be  the gaseous progenitors of current galaxies
(see Wolfe 1995), 
provide an independent perspective for studying these
events  because (a) they occur in 
objects comprising the bulk of the
galaxy population at high redshifts, and 
(b) the ranking of redshifts
yields an unambiguous time sequence.
Detected in the redshift
interval $z$ = [0,4.5] the damped {\lya} systems trace the evolution of neutral 
gas in galaxies from their protogalactic phase to the present.
However, the chemical properties of the gas may be incompatible with those
of existing stellar populations.
At $z$ = [1.6,4.5] the metallicity of the gas is
low compared to the thin disk metallicity (Pettini {\etal} 1997) indicating
that stars in the thin disk do not arise directly from  high-$z$ damped
systems (Lanzetta {\etal} 1995). The possible enhancement of alpha-rich elements
suggests the gas gives rise to halo stars 
(Lu {\etal} 1996), but the kinematics of the gas 
are inconsistent with this hypothesis (Prochaska \& Wolfe 1997). More
recently Pettini {\etal} (1997) argued that the metallicities 
of the damped systems are too low to explain the thick disk (Gilmore {\etal} 1989;
Carney {\etal} 1996).

In this letter we reconsider the scenario in which star formation in
damped {\lya} systems results in the formation of the thick
disk. Combining metallicities and column densities with new kinematic data  obtained with the Keck I 10 m
telescope we suggest a plausible
scenario in which the thick disk forms out of damped {\lya} gas. 
 
\section{COSMIC METALLICITY DEPENDENCE ON BARYON DENSITY}
 
We wish to find whether 
the mass content and metal abundances of
gas in damped {\lya} systems
can account for the
mass density and metallicities of thick stellar disks.
Define the cosmic metallicity $<{\cal Z}>$ $\equiv$
$\Omega_{metals}$/$\Omega_{g}$ where  $\Omega_{metals}$ and $\Omega_{g}$
are the comoving densities of metals and neutral gas 
in damped {\lya} systems (Lanzetta {\etal} 1995). Let the number of damped systems
in the metallicity and column-density intervals
($Z^{\prime}$,$Z^{\prime}$$+$d$Z^{\prime}$) and
($N$,$N$$+$d$N$)
be given by $h(Z^{\prime},N)dZ^{\prime}$d$N$. The
latter is
related to the frequency distribution of H I column
densities by
$f(N)$=$\int h(Z^{\prime},N)dZ^{\prime} $
(Lanzetta {\etal} 1995), and the frequency distribution of
metallicities by $g(Z^{\prime})$=$\int h(Z^{\prime},N)dN $.
Suppose $h(Z^{\prime},N)$ spans the metallicity
interval $Z^{\prime}$ = [$Z_{min},Z_{max}$] and column-density
interval $N$ = [$N_{min},N_{max}$],
and $\Omega_{g}(Z)$ is the density of damped
{\lya} baryons in the metal-rich subinterval $Z^{\prime}$ = [$Z,Z_{max}$].
Then $\Omega_{g}(Z)$ and the corresponding $<{\cal Z}(Z)>$  are given by

\begin{equation}
\Omega_{g}(Z) = \Omega_{g} \times
{\intl_{Z_{max}}^{Z} dZ^{\prime} \intl_{N_{min}}^{N_{max}} 
dN N \, h(Z^{\prime},N) \over
\intl_{Z_{max}}^{Z_{min}} dZ^{\prime}
\intl_{N_{min}}^{N_{max}} dN N \, h(Z^{\prime},N)} \cmma
\end{equation}

\begin{equation}
{<{\cal Z}(Z)>} = {\intl_{Z_{max}}^{Z} dZ^{\prime}
\intl_{N_{min}}^{N_{max}} dN \, Z^{\prime} \, N \, h(Z^{\prime},N)
\over \intl_{Z_{max}}^{Z} dZ^{\prime} \intl_{N_{min}}^{N_{max}} 
dN N \, h(Z^{\prime},N)} \cmma
\end{equation}

\noindent where the order of $Z$ integration is reversed.
In the discrete limit $h(Z^{\prime},N)$ = $\smm_{i}
\delta(Z^{\prime}-Z_{i})\delta(N-N_{i})$,
where the sum extends over all the $N_{i}, Z^{\prime}_{i}$ pairs in the sample. As a result

\begin{equation}
{\Omega_{k}} = {\Omega_{g}} \, \times \, {\smm_{i=1}^{k} N_{i}
\over{\smm_{j=1}^{i_{min}}N_{j}}} \cmma \;\;\;
<{\cal Z}_{k}> = {{\smm_{i=1}^{k}N_{i}{\times}Z^{\prime}_{i}} 
\over{\smm_{j=1}^{k}N_{j}}} \cmma
\end{equation}

\noindent where  the indices $i$=1, $k$, and $i_{min}$  correspond to $Z_{max}$, $Z$,
and $Z_{min}$.
Because the sums in eq. (3) are over
an array of damped {\lya} gas layers ordered according to
{\em decreasing} metallicity, 
$<{\cal Z}(Z)>$ decreases
with {\em decreasing} $Z$ while ${\Omega_{g}(Z)}$ increases.
We can determine ${\Omega_{g}(
Z)}$ corresponding
to the mean metallicity of the thick disk,
provided the latter
is less than
$Z_{max}$.

To determine {\cosmZ} as a function of ${\Omega_{g}(Z)}$ we turn to 
the [Zn/H], {\colH} pairs that Pettini {\etal} (1997) acquired for 34
damped systems, where $Z$=$Z_{\odot}$10$^{[{\rm Zn/H}]}$. We focus on Zn rather than Fe as a metallicity indicator, because
this is the largest recorded sample of damped {\lya} metal abundances, and Zn is 
less depleted than Fe by dust which may be present (Fall \& Pei 1995).
We select
the 27 pairs 
in the redshift range $z$ =[1.6,3.0].
Systems with $z$ $>$ 3.0 are excluded, since the metallicities
in this redshift range are systematically lower than those of the thick disk. 
The sample
comprises 16 systems with detected [Zn/H] and 11 with upper limits. 
Two of the detections,
which come from our Keck HIRES observations, replace 
the upper limits of Pettini {\etal} (1997).

We used
equation (3) to determine the points
in Figure 1 which plots [$<$Zn/H$>$] ($\equiv$ log(${\cal Z}_{k}/Z_{\odot})$) vs ${\Omega}_{k}$
for $k$ =
[1,$i_{min}$]. We let $\Omega_{g}$ = 0.003, the value inferred
by Storrie-Lombardi \& Wolfe (1997) in the redshift interval $z$ = [1.8,3.5]
for $H_{0}$ = 50 {\kms} Mpc$^{-1}$ (which is adopted throughout this paper).
The 
circles were computed
by letting the upper limits equal the true values of [Zn/H]. In this case {$\Omega_{g}(Z)$} = 0.0004
when  [$<$Zn/H$>$] = $-$ 0.6, the mean metallicity of the thick disk (Carney  {\etal} 1996).
The triangles and squares
were computed by equating the upper limits minus 0.5 and 1.0 with the true values of [Zn/H].
In both cases {$\Omega_{g}(Z)$} $\approx$  0.0003 when [$<$Zn/H$>$] = $-$ 0.6, indicating the
result is robust. 
Assuming bulges and disks contribute equally
to the density of visible matter (Schechter \& Dressler 1987),

\break

\begin{figure}
\begin{center}
\includegraphics[height=4.0in, width=4.0in]{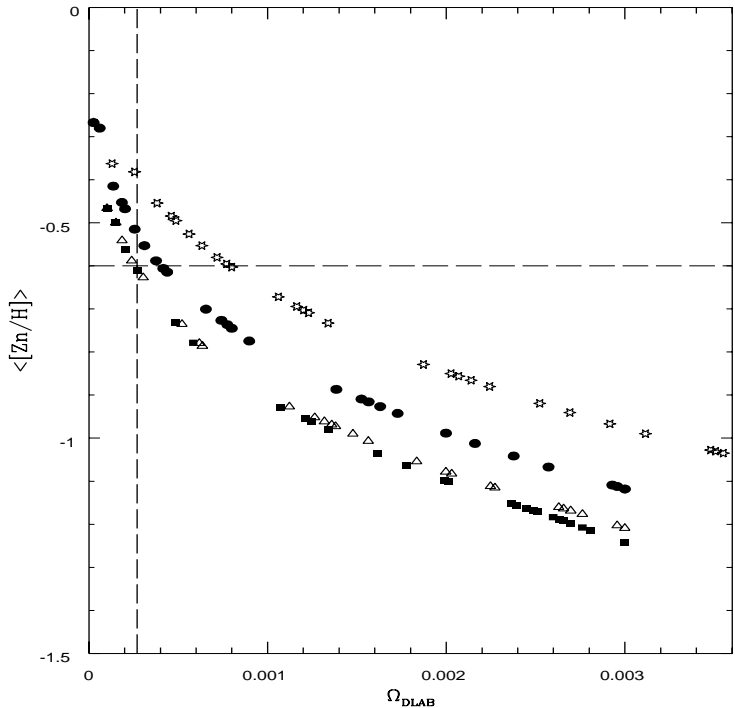}
\end{center}
\caption {Log of mean cosmic metallicity vs the comoving
density
of damped {\lya} baryons, for damped systems with metallicities $Z =[Z,Z_{max}]$.
Circles computed assuming upper limits equal true values of [Zn/H]. Triangles
and squares computed assuming upper limits minus 0.5 and 1.0 equal true values
of [Zn/H]. In latter cases all baryons are assumed to be gas. Thus eq. (3) is
used to compute
$\Omega_{DLAB}(Z)$ = $\Omega_{k}$ and  
[$<$Zn/H$>$] = log($<{\cal Z}_{k}>$/$Z_{\odot}$).  
Stars include correction for the presence of baryons in stars. In this case eqs. (4)
$-$(6)  are used to compute $\Omega_{DLAB}(Z)$ and [$<$Zn/H$>$]. 
Vertical and horizontal dashed lines correspond to 
cosmic density and metallicity of the thick stellar disk.}
\label{sptra}
\end{figure}

\noindent which is given by 0.0054 (Gnedin \& Ostriker 1992),
and that the mass of the thick disk is 0.1 times that of the thin
disk (Majewski 1993) we find the thick
disk mass density, 
$\Omega_{thick}$ = 0.00027. Although the
error bars associated with $\Omega_{thick}$ are of order 50$\%$, it is
reasonable to conclude that the damped {\lya}
systems contain sufficient baryons to account for the masses of thick 
stellar disks (see Figure 1).

\section{KINEMATICS, METALLICITIES, AND STARS}
\label{mdls}

To learn more about the metal-rich damped systems 
we turn to the kinematics of the gas. Analysis of
the velocity profiles of weak metal lines in over 30 damped {\lya} systems shows  
the frequency distribution of profile velocity widths, $\Delta v$, and other
statistics that test for asymmetries exhibited by the profiles are consistent with 
absorption by thick disks with rotation speeds $v_{rot}$ $\approx$
250 km s$^{-1}$(Prochaska \& Wolfe 1997). The CDM simulation 
of Haehnelt {\etal} (1997), in which infall, random motions, and rotation of
protogalactic clumps contribute equally to
$\Delta v$, 
may be likewise consistent. 
Here we focus on rotating disks. 

\begin{figure}
\centering
\includegraphics[height=5.0in, width=5.0in]{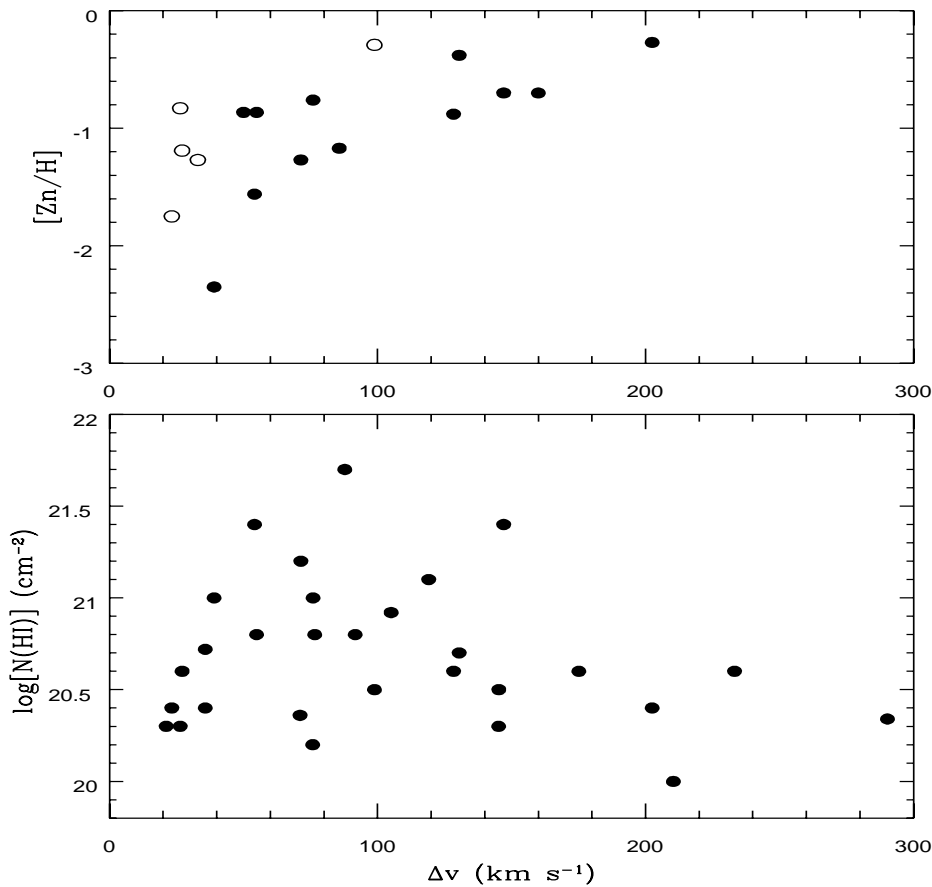}
\caption{(a) metallicity vs $\Delta v$ for
damped systems with $z$ = [1.6,3.0]. Filled and empty circles correspond to
detections and upper limits. (b) log{\colH} vs
$\Delta v$.} 
\label{obstau}
\end{figure}

Figure 2a plots 17 [Zn/H], $\Delta v$ pairs drawn from our kinematic
sample with $z$ =[1.6,3.0].
The figure shows
that  systems with high $\Delta v$ and low metallicity
are {\em not} detected in this redshift range.
Specifically metallicities [Zn/H] $<$ $-1.0$  are absent in all 5 systems with
$\Delta v$ $>$ 120 {\kms}, but present in 7 out of 12 systems with 
$\Delta v$ $<$ 120 {\kms}.
The effect 
is real and is not an artifact due to observational selection,
since systems with high $\Delta v$ and low metallicity are
detected at $z$ $>$ 3.0. Nor is dust likely to be a contributing factor,
since dust would remove metal-rich rather than the metal-poor
systems missing from Figure 2a. 
The reality of this effect is
further supported by its presence in 
a [Fe/H] vs $\Delta v$ diagram. 
We also find possible evidence for a
correlation
between [Zn/H] and $\Delta v$ exceeding 3.5$\sigma$ significance when
the true [Zn/H] equal the upper limits minus 1.0.

The systematic pattern in Figure 2a can be explained by  
negative radial gradients in metallicity.  
Monte Carlo simulations of absorption profiles produced by
sightlines penetrating randomly oriented disks 
indicate that a necessary condition for large
$\Delta v$ is a small impact 
parameter.  This is evident in
Figure 3 which plots the distribution of impact parameters, $b$
(where $b$ is in units of radial scale length, $R_{d}$, of an assumed exponential
gas distribution),
resulting from simulating identical exponential
disks with rotation speed $v_{rot}$ = 250 {\kms} and vertical scale-height
$h$ = 0.3$R_{d}$. Whereas 86$\%$ of impacts 
leading to {\Dv} $>$ 120 {\kms} 
are confined to $b$ $<$ 1, 1$\%$ are at  $b$ $>$ 2. 
Therefore, the absence of low metallicities at $\Delta v$ $>$ 120 {\kms} requires
high element abundances at small radii.
On the other hand 26$\%$ of impacts 
leading to $\Delta v$ $<$ 120 {\kms} are at $b$ $<$ 1,
while 16$\%$ are at $b$ $>$ 2.
The wide range of impact parameters 
can explain the broader distribution of metallicities at $\Delta v$ $<$ 120 {\kms}, if
impacts at large $b$ yield
low metallicities, i.e., if metallicity decreases with radius.
None of these results changes significantly when we use
a more realistic model in which $v_{rot}$ is drawn
from a distribution of rotation speeds characterizing 	present-day
spiral galaxies.

\begin{figure}
\begin{center}
\includegraphics[height=4.0in, width=4.0in]{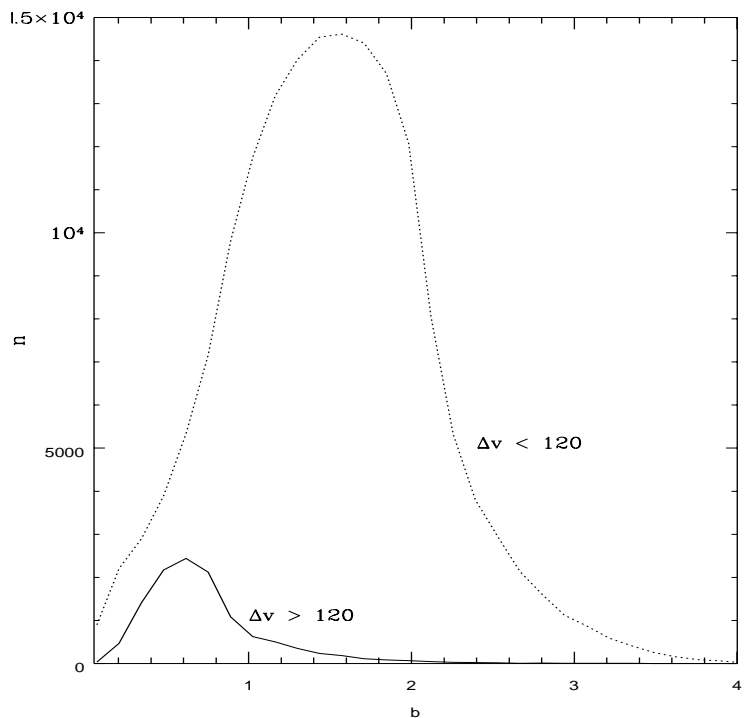}
\end{center}
\caption{Distribution of impact parameters resulting from the numerical
simulation described in text. Solid curve corresponds to impacts leading to {\Dv} $>$ 120 {\kms}, and
dotted curve to {\Dv} $<$ 120 {\kms}. Impact parameters in units of radial
scale-length, $R_{d}$.}  
\label{mlos}
\end{figure}

Damped {\lya} systems with large $\Delta v$ also exhibit
systematically lower {\colH}. This is shown in Figure 2b which
plots log{\colH} vs $\Delta v$ for 29 damped {\lya} systems
drawn from our kinematic sample. 
Whereas 1 out of 10 systems with $\Delta v$
$>$ 120 {\kms} has
log{\colH} $>$ 20.6, 12 out of 19 systems with $\Delta v$ $<$ 120 {\kms}
have log{\colH} $>$ 20.6.
Figure 2b includes systems with $z$ $>$ 3.0, 
because the effect is independent of redshift.
Suppose the gas distribution has a central hole. 
At high $\Delta v$ the impact parameters are so small
that the sightlines encounter the low column densities
present at small radii. A wider range of {\colH} occur at
low {\Dv}, because the sightlines sample a broader range
of impact parameters.
Preliminary results from simulations
with central holes are  in better agreement with the log{\colH}
vs $\Delta v$ 
data than standard exponential disks.

Deficiency of neutral gas occurs often in the central regions of spiral galaxies 
(Broeils \& van Woerden 1994), the same regions where
enhancements in metallicity are also common (Edmunds \& Pagel 1984).
Thus, there is empirical support for the idea that damped
{\lya} systems comprise gaseous disks with central holes and negative
radial gradients in metallicity, if they evolve into
current spirals. The increased metallicity is a signature of enhanced star
formation which also helps to explain the deficit of gas, either through
direct gas consumption or the loss of gas through
energetic outflows from supernovae. In either case a significant fraction
of baryons in the metal-rich damped systems may be 
locked up in stars. As a result the expression for cosmic metallicity
in eq. (3) will underestimate the contribution from the gas-poor
metal-rich systems.

When stars are present, ${\Omega_{k}}$ and $<{\cal Z}_{k}>$  
are given by

\begin{equation}
{\Omega_{k}} = {\Omega_{g}}{\times}{{\smm_{i=1}^{k}(N_{i}+N_{i}^{s})}
\over{\smm_{j=1}^{i_{min}}N_{j}}} \cmma \;\;
<{\cal Z}_{k}> = {\smm_{i=1}^{k}(N_{i}{\times}Z^{\prime}_{i}+N_{i}^{s}
{\times}Z_{i}^{s}) \over \smm_{j=1}^{k}(N_{j}+N_{j}^{s})} \cmma
\end{equation}

\noindent where ${\Omega_{k}}$ is the comoving mass density of baryons in stars plus gas.
Because $N_{i}^{s}$ and $Z_{i}^{s}$ are the column density and metallicity of matter in
stars, $<{\cal Z}_{k}>$ is the comoving mass density of metals in stars plus gas divided
by {$\Omega_{k}$}.
Although $<{\cal Z}_{k}>$ in eq. (4) differs from the standard definition for metallicity,
it is the appropriate quantity, 
because metals in stars as well as gas in damped {\lya} systems supply metals to stars comprising the
current thick disk. And the thick disk metallicity is inferred solely from stars.

To solve eq. (4) we first adopt the chemical evolution model of Larson (1972) to
compute the fraction of baryons in stars. 
The model assumes the
star formation rate is balanced by the rate of mass infall to the disk, and as a result
the gas content of the galaxy does not change. This agrees with the observed
constancy
of $\Omega_{g}(z)$ in the redshift range $z$ = [1.6,3.3] (Storrie-Lombardi
\& Wolfe 1997) in which stars of the thick disk are assumed to form. We have

\begin{equation}
N_{i}^{s}/N_{i} = {\rm ln}\Biggl[{{y+Z_{f}-Z_{init}} \over{y+Z_{f}-Z^{\prime}_{i}}}\Biggr] \cmma
\end{equation}

\noindent where $y$ is the chemical yield, and $Z_{f}$ and $Z_{init}$ are the metallicities
of the infalling material and of the  `initial' disk at $z$ $>$ 3.0. We determine
$Z_{i}^{s}$ from the constraint

\begin{equation}
Z_{i}^{s}{\times}N_{i}^{s}+(Z^{\prime}_{i}-Z_{init}){\times}N_{i} = y{\times}N_{i}^{s} \cmma
\label{Ncolm}
\end{equation}

\noindent (see Tinsley 1980).
We combined eqs. (4)$-$(6) to compute log(${\cal Z}_{k}/Z_{\odot}$)
versus ${\Omega_{k}}$ in the presence of stars. The solution, shown as stars
in Figure 1, was computed assuming $y = 0.5Z_{\odot}$, $Z_{init} = Z_{f} = 0.01Z_{\odot}$.
We estimated $Z_{init}$ and $Z_{f}$ from the lowest metallicities found
for damped {\lya} systems at $z$ $>$ 3.0, and $y$ from standard models (Tinsley 1980).
In this case ${\Omega_{k}}$ = 0.0008, i.e., 3{$\Omega_{thick}$,
when log(${\cal Z}_{k}/Z_{\odot}$) = $-$ 0.6.
The increase in ${\Omega_{k}}$ at the metallicity
of the thick disk results from the significant stellar corrections in baryonic mass 
for the metal rich damped {\lya} systems.

\section{DISCUSSION AND CONCLUSIONS}

Our results suggest that contrary to previous claims 
(Pettini {\etal} 1997) the damped {\lya} systems contain more than 
enough baryons at
suitable metallicity and rotation speed to form thick stellar disks in spiral galaxies.
The kinematic/metallicity data further imply (a) stars in the thick disk
form in the inner metal-rich regions of rapidly rotating gaseous disks, and (b)
these stars may
dominate the baryonic content of the most metal rich damped systems. We conjecture
that vertical contraction of the metal-rich, thick gaseous disk
leads to the formation
of the inner thin disk. The metal-poor gas of the outer disk, 
i.e.,  the
$\sim$ 90 $\%$ of the damped {\lya} baryons that remain in gas
after the formation of the thick disk, could supply
the remaining thin-disk mass 
through radial contraction driven by
angular momentum transport
mediated by high-amplitude spiral density waves (Roberts \& Shu 1972).
The metallicity of this
gas may increase as a result of mass loss 
by stars in the inner regions. 

By contrast stars forming
in the protogalactic clumps considered by Haehnelt {\etal} (1997)
end up in
bulges and halos rather than
rotationally supported structures, because the 
motions of the merging clumps are 
not dominated by rotation.
Formation of the thick disk occurs at $z$ $<$ 1, after merging ceases, and
when the mean metallicity equals $-$0.6.
But the age of the thick disk is unlikely to be
less than 12 Gyr (Carney 1997) which
exceeds the lookback time to $z$ = 1 in all $\Lambda$ = 0 cosmologies
and spatially flat
cosmologies in which $\Omega_{\Lambda}$ $<$ 0.8, when $H_{0}$ $>$ 50 {\kms} Mpc$^{-1}$. 
Edge-leading  asymmetries of the velocity profiles
arise in this model because 
clumps with the highest volume and column density move
fastest with respect to the surrounding gas. This may occur because
ram pressure
deceleration by ambient gas is less effective in decelerating
denser clumps. The predicted
correlation between {\colH} and {\Dv} is in conflict with
the detection of systems having high
{\Dv} (i.e., $>$ 120 {\kms}) and low log{\colH} ($<$ 20.6)(see Figure 2b).

\vskip 0.5in

We wish to thank W. L. W. Sargent and L. Lu for providing HIRES data prior
to publication, and B. Carney,  A. Loeb, and N. Reid  for valuable discussions.
We also thank the referee, K. Lanzetta for valuable comments that improved 
the presentation of this paper. AMW was a guest of the Center for Particle
Astrophysics at U. C. Berkeley during the inception of this work and he
thanks M. Davis, B. Sadoulet, and J. Silk for their kind hospitality.
The authors 
were partially supported by 
NASA grant NAGW-2119 and NSF grant AST 86-9420443.

\end{document}